\documentstyle[11pt,epsf]{article}
\title{Magnetoelectric subbands and eigenstates in the presence of Rashba and 
Dresselhaus 
spin-orbit interactions in a quantum wire}
\author{S. Bandyopadhyay$\thanks{E-mail: sbandy@vcu.edu}$ and S. Pramanik \\
Department of Electrical Engineering \\
Virginia Commonwealth University \\
Richmond, VA 23284, USA \\
\\M. Cahay \\
Department of Electrical and Computer Engineering and Computer Science \\
University of Cincinnati \\
Cincinnati, OH 45221, USA}
\date{}

\begin{document}

\maketitle

\begin{abstract}
We derive the eigenenergies and eigenstates of electrons  in a quantum wire 
subjected to an external magnetic field. These are calculated in the presence of 
spin orbit interactions  arising from the Rashba 
(structural inversion asymmetry) and Dresselhaus (bulk inversion asymmetry) 
effects. We consider  three  cases: the external magnetic 
field is oriented (i) along the axis of the wire, (ii)
 perpendicular to the axis but parallel to the electric field associated with 
structural inversion asymmetry (Rashba effect), and (iii) 
perpendicular to the axis as well as the electric field.  In all cases, the 
dispersions of the eigenenergies 
are non-parabolic and the subbands do not have a fixed spin quantization axis 
(meaning that the spin polarization of the electron is wavevector dependent). 
Except in the 
second case, the dispersion diagrams are also, in general, asymmetric about the 
energy axis.
\end{abstract}

\newpage

\section{Introduction}

Consider a semiconductor quantum wire defined by split gates  on a two 
dimensional electron gas confined at a heterointerface. We will derive the
eigenstates and the dispersion relations of the eigenenergies in this  wire when 
it is placed in an 
external 
magnetic field of flux density $B$ (Fig. 1).

Referring to Fig. 1, the potential in the y-direction (perpendicular to the 
heterointerface) is 
approximately triangular which leads to a 
structural inversion asymmetry and an associated Rashba spin orbit interaction 
\cite{rashba}. Additionally, if the material has bulk inversion asymmetry (e.g. 
InAs), then there will be a Dresselhaus interaction \cite{dresselhaus} which is 
important at large wavevectors. The potential in the z-direction will be 
approximately parabolic and hence symmetric 
about a center under inversion. Thus, there is no significant overall 
Rashba interaction associated with the z-directed electric field.

We will consider three cases corresponding to the external magnetic field being 
oriented 
along three possible coordinate axes -- $x$, $y$ and $z$.

\section{Magnetic field directed along the wire axis (i.e. x-axis)}

When the magnetic field is oriented along the axis of the wire, the 
effective mass Hamiltonian for the wire, in the Landau gauge 
${\bf A}$ = (0, $-Bz$, 0),  can be written as
\begin{eqnarray}
H & = & (p_x^2 + p_y^2 + p_z^2)/(2m^*) + (e B z p_y)/m^* + (e^2 B^2 z^2)/(2m^*) 
- (g/2) \mu_B B \sigma_x  \nonumber \\
& & + V(y) + V(z) + \nu [\sigma_x \kappa_x + \sigma_y \kappa_y + \sigma_z 
\kappa_z ] + \eta [ 
(p_x/\hbar) \sigma_z - (p_z/\hbar) \sigma_x ]
\end{eqnarray}
where $g$ is the Land\`e g-factor, $\mu_B$ is the Bohr magneton, $V(y)$ and 
$V(z)$ are the confining potentials along the y- and z-directions, $\sigma$-s 
are the Pauli spin matrices, $\nu$ is the strength of the Dresselhaus 
spin-orbit interaction  and $\eta$ is the strength of the 
Rashba spin-orbit 
interaction. 

\begin{figure}[t]
\epsfxsize=4.3in
\epsfysize=4.3in
\centerline{\epsffile{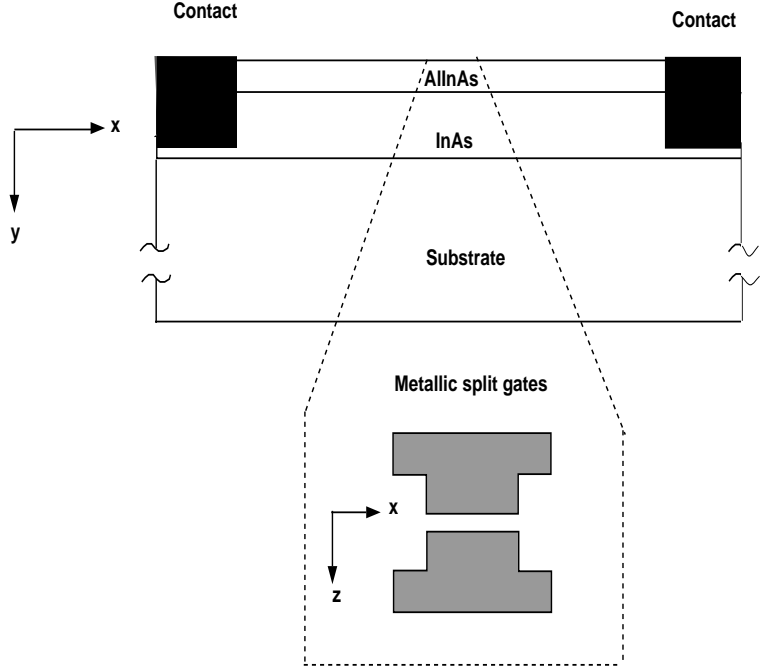}}
\caption[]{\small A generic semiconductor quantum wire defined by split gates at 
a heterostructure interface.}
\end{figure}

The quantities 
$\kappa$ are defined in ref. \cite{das}.
We will assume that the wire is narrow enough and the temperature is low enough 
that 
only the lowest magneto-electric subband is occupied. Since the Hamiltonian is 
invariant in the $x$-coordinate, the wavevector $k_x$ is a good quantum number 
and the eigenstates are plane waves traveling in the x-direction. Accordingly,  
the spin Hamiltonian (spatial operators are replaced by their expected values)
simplifies to  
\begin{equation}
H = (\hbar^2 k_x^2)/(2 m^*) + E_0 +  (\alpha_1 k_x -  
\beta) 
\sigma_x + \eta k_x \sigma_z
\end{equation}
where $E_0$ is the energy of the lowest magneto-electric subband,   $\alpha_1(B) 
=  
\nu [ <k_y^2> - <k_z^2>
+ (e^2 B^2 <z^2>/\hbar^2)]$, $\psi(z)$ is the z-component of the wavefunction, 
$\phi(y)$ is the 
y-component of the wavefunction,  $<k_y^2>$ = 
$(1/\hbar^2)<\phi(y)|-(\partial^2/\partial y^2) |\phi(y)>$, $<k_z^2>$ = 
$(1/\hbar^2)<\psi(z)|-(\partial^2/\partial z^2) |\psi(z)>$, and $\beta = (g/2) 
\mu_B B$.

Since the potential $V(z)$ is parabolic ($V(z) = (1/2)m^* \omega_0^2 z^2$), it 
is 
easy to show that $<k_z^2> = m^* \omega/(2 \hbar)$ and $<z^2> = \hbar/( 2 m^* 
\omega)$ where $\omega^2 = \omega_0^2 + \omega_c^2$ and $\omega_c$ is the 
cyclotron frequency ($\omega_c = e B/m^*$). Furthermore, $E_0$ = $(1/2) \hbar 
\omega + E_{\Delta}$ where $E_{\Delta}$ is the energy of 
the lowest subband in the triangular well $V(y)$.

Diagonalizing this Hamiltonian in a truncated Hilbert space spanning the two 
spin resolved states in the lowest subband yields the eigenenergies
\begin{equation}
E^{(1)}_{\pm} = {{\hbar^2 k_x^2}\over{2 m^*}}  + E_0 \pm 
\sqrt{\left ( \eta^2 + \alpha_1^2 \right ) \left ( k_x - {{\alpha_1 
\beta}\over{\eta^2 + \alpha_1^2}} \right)^2 + {{\eta^2}\over{\eta^2 + 
\alpha_1^2}} 
\beta^2} 
\label{eigenenergy}
\end{equation}
and the corresponding eigenstates
\begin{eqnarray}
{\Psi^{(1)}}_{+}(B, x) =
\left [ \begin{array}{c}
             cos(\theta_{k_x})\\
             sin(\theta_{k_x}) \\
             \end{array}   \right ]
             e^{i  k_x x}
             ~~~~~~~~~
{\Psi^{(1)}}_{-}(B, x) =
 \left [ \begin{array}{c}
              sin(\theta_{k_x})\\
               - cos(\theta_{k_x})\\
             \end{array}   \right ]
             e^{i  k_x x}
\label{eigenstate}
\end{eqnarray}
where $\theta_{k_x}$ = $(1/2) arctan [(\alpha_1 k_x - \beta)/\eta k_x]$. 

The dispersion relations given by Equation (\ref{eigenenergy}) are plotted in 
Fig. 2 for the case $\eta$ = $\alpha_1$. Note that the dispersions are clearly 
nonparabolic and {\it asymmetric} about the energy axis. More importantly, 
note that the eigenspinors given in Equation 
(\ref{eigenstate}) are functions of $k_x$ because $\theta_{k_x}$ depends on 
$k_x$. Therefore, the eigenspinors are not fixed in any subband, but change with 
$k_x$. In other words, neither subband has a definite spin quantization axis and 
the orientation of the spin vector of an electron in either subband depends 
on the wavevector. Consequently, it is always possible to find two states in the 
two subbands with non-orthogonal spins. Any non-magnetic scatterer (impurity, 
phonon, etc.) can then couple these two states and cause a spin-relaxing 
scattering event. 

\begin{figure}
\epsfxsize=4.3in
\epsfysize=4.3in
\centerline{\epsffile{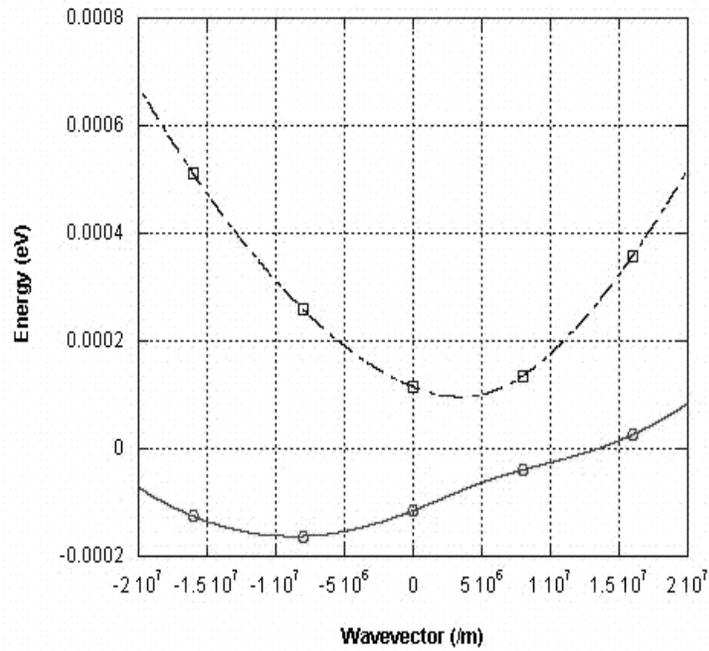}}
\caption[]{\small Energy versus wavevector $k_x$ in a quantum wire where we have 
assumed $\alpha_1$ = $\eta$ = 10$^{-11}$ eV-m, the magnetic flux density = 1 
Tesla, the effective mass $m^*$ is 0.05 times the free electron mass and the 
Land\'e g-factor = 4. This corresponds to $\beta$ = 0.115 meV. The magnetic 
field is in the x-direction (along the wire).}
\end{figure}

\subsection{Spin components}

The spin components along the $x$, $y$ and $z$ directions are given by
\begin{equation}
S^{(1) \pm}_m = [\Psi^{(1)}_{\pm}]^{\dag}[\sigma_m][\Psi^{(1)}_{\pm}] ~~~ m = 
x,y,z
\end{equation}
This yields $S^{(1) \pm}_x (k_x) = \pm sin(2 \theta_{k_x})$, $S^{(1) \pm}_y$ = 0 
and 
$S^{(1) \pm}_z (k_x) = \pm cos(2 \theta_{k_x})$

It is easy to see that, in general, $S^{(1)}_x(k_x) \neq S^{(1)}_x(-k_x)$ and 
$S^{(1)}_z(k_x) \neq S^{(1)}_z(-k_x)$. However, if the Dresselhaus interaction 
vanishes ($\alpha_1$ = 0), 
then $S^{(1)}_x(k_x) = - S^{(1)}_x(-k_x)$ and $S^{(1)}_z(k_x) = 
S^{(1)}_z(-k_x)$. This {\it does not} always mean that, in the absence of 
Dresselhaus interaction, oppositely traveling electrons have anti-parallel spin 
components along the wire axis, and parallel spin components in the z-direction. 
 The reason being that the velocity operator is {\it not} $p/m^*$ and hence 
electrons with opposite signs of wavevector $k_x$ need not have opposite signs 
of velocity. 

\section{Magnetic field perpendicular to wire axis and along the electric field 
causing Rashba effect (i.e. along y-axis)}

With this orientation of the magnetic field, we will use the Landau gauge 
${\bf A}$ =  ($Bz$, 0, 0), and the Hamiltonian for the wire can be written as
\begin{eqnarray}
H & = & (p_x^2 + p_y^2 + p_z^2)/(2m^*) - (e B z p_x)/m^* + (e^2 B^2 z^2)/(2m^*) 
- (g/2) \mu_B B \sigma_y  \nonumber \\
& & + V(y) + V(z) + \nu [\sigma_x \kappa_x + \sigma_y \kappa_y + \sigma_z 
\kappa_z ] + 
(\eta/\hbar) [ 
(p_x + e B z) \sigma_z - p_z \sigma_x ]
\end{eqnarray}

Using the definition of the 
$\kappa$-s, and noting that the expected values $<z>$ = $<p_z>$ = 0,
the spin Hamiltonian can be
simplified to  
\begin{equation}
H = (\hbar^2 k_x^2)/(2 m^*) + E_0 +  \alpha_2 k_x 
\sigma_x - \beta \sigma_y + \eta k_x \sigma_z
\end{equation}
where  $\alpha_2 
(B) =  
\nu [ <k_y^2> - <k_z^2>
]$.

Again, diagonalizing this Hamiltonian in a truncated Hilbert space spanning the 
two 
spin resolved states in the lowest subband yields the eigenenergies
\begin{equation}
E^{(2)}_{\pm} = {{\hbar^2 k_x^2}\over{2 m^*}}  + E_0 \pm 
\sqrt{\left ( \eta^2 + \alpha_2^2 \right ) k_x^2 +  
\beta^2} 
\label{eigenenergy1}
\end{equation}
and the corresponding eigenstates
\begin{eqnarray}
{\Psi^{(2)}}_{+}(B, x) =
\left [ \begin{array}{c}
             cos(\theta'_{k_x})\\
             sin(\theta'_{k_x})e^{-i \phi_{k_x}} \\
             \end{array}   \right ]
             e^{i  k_x x}
             ; ~~
{\Psi^{(2)}}_{-}(B, x) =
 \left [ \begin{array}{c}
              sin(\theta'_{k_x})\\
              - cos(\theta'_{k_x})e^{-i \phi_{k_x}}\\
             \end{array}   \right ]
             e^{i  k_x x}
\label{eigenstate1}
\end{eqnarray}
where $\theta'_{k_x}$ = $(1/2) arctan [\sqrt{(\alpha_2 k_x)^2 + \beta^2)}/\eta 
k_x]$ and $\phi_{k_x}$ = $ arctan [\beta/\alpha_2 k_x]$. 

In Fig. 3, we plot the dispersion relations for the case $\eta$ = $\alpha_2$. 
Note that the dispersions are again clearly nonparabolic, but this time they are 
symmetric about the energy axis. The eigenspinors are functions of $k_x$ because 
$\theta'_{k_x}$ and $\phi_{k_x}$ depend on $k_x$. Therefore, the eigenspinors 
are again not fixed in any subband, but change with $k_x$. Neither subband has a 
definite spin quantization axis and the spin of an electron in either subband 
depends 
on the wavevector. Consequently, it is again always possible to find two states 
in the 
two subbands having non-orthogonal spins and a non-magnetic scatterer  can 
couple these two states causing a spin-relaxing 
scattering event.

\begin{figure}
\epsfxsize=4.3in
\epsfysize=4.3in
\centerline{\epsffile{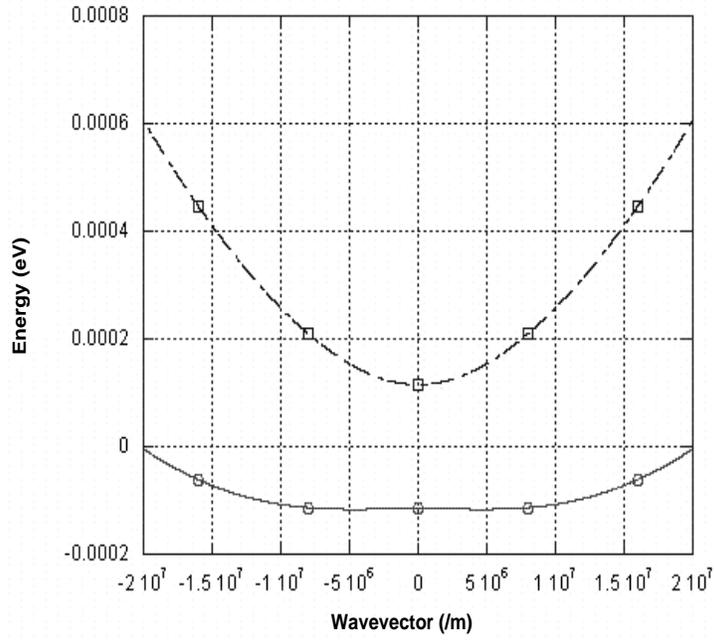}}
\caption[]{\small Energy versus wavevector $k_x$ in a quantum wire where the 
magnetic flux density is along the y-direction. We have assumed $\alpha_2$ = 
$\eta$ = 10$^{-11}$ eV-m. All other parameters are the same as in Fig. 2.}
\end{figure}

\subsection{Spin components}

The spin components along the $x$, $y$ and $z$ directions are given by
\begin{equation}
S^{(2) \pm}_m = [\Psi^{(2)}_{\pm}]^{\dag}[\sigma_m][\Psi^{(2)}_{\pm}] ~~~ m = 
x,y,z
\end{equation}
This yields $S^{(2) \pm}_x (k_x) = \pm sin(2 \theta'_{k_x}) cos (\phi_{k_x})$, 
$S^{(2) \pm}_y$ = 0 and 
$S^{(2) \pm}_z (k_x) = \pm cos(2 \theta'_{k_x})$.

Again, $S^{(2)}_x(k_x) \neq S^{(2)}_x(-k_x)$ and $S^{(2)}_z(k_x) \neq 
S^{(2)}_z(-k_x)$. However, if the Dresselhaus interaction vanishes ($\alpha_2$ = 
0), 
then $S^{(2)}_x(k_x) = - S^{(2)}_x(-k_x)$ and $S^{(2)}_z(k_x) = 
S^{(2)}_z(-k_x)$.

\section{Magnetic field perpendicular to wire axis and  the electric field 
causing Rashba effect (i.e. along z-axis)}

In this case, using the Landau gauge 
${\bf A}$ =  ($-B(y-y_0)$, 0, 0), the Hamiltonian for the wire can be written as
\begin{eqnarray}
H & = & (p_x^2 + p_y^2 + p_z^2)/(2m^*) + (e B y p_x)/m^* + (e^2 B^2 y^2)/(2m^*) 
- (g/2) \mu_B B \sigma_z  \nonumber \\
& & + V(y) + V(z) + \nu [\sigma_x \kappa_x + \sigma_y \kappa_y + \sigma_z 
\kappa_z ] + 
(\eta/\hbar) [ 
(p_x + e B y) \sigma_z - p_z \sigma_x ]
\end{eqnarray}
where $y_0$ is a gauge constant.

Using the definition of the 
$\kappa$-s, 
the spin Hamiltonian can be
simplified to  
\begin{eqnarray}
H & = & (\hbar^2 k_x^2)/(2 m^*) + (eB \hbar/m^*) (<y> - y_0)k_x + E'_0 +  
(\alpha_3 k_x + \Delta)
\sigma_x \nonumber \\
&& - \beta \sigma_z + \eta (k_x + eB (<y> - y_0)/\hbar) \sigma_z
\end{eqnarray}
where $E'_0$ is the energy of the lowest magneto-electric subband,   $\alpha_3 =  
\nu [ <k_y^2> - <{k'}_z^2>]$, and $\Delta$ = $(\nu e B/\hbar) I(B)$ where $I(B)$ 
= 
$\int_{\infty}^{\infty} (y - y_0) (\partial \zeta(y, B)/\partial y)^2 dy$, 
$\zeta(y, B)$ 
is the y-component of the wavefunction and $<y>$ =  $\int_{\infty}^{\infty} 
|\zeta(y, B)|^2 y dy$. We choose the gauge constant $y_0$ such that $<y>$ = 
$y_0$. This simplifies the Hamiltonian to 

\begin{equation}
H = (\hbar^2 k_x^2)/(2 m^*) +  + E'_0 +  (\alpha_3 k_x + \Delta)
\sigma_x - \beta \sigma_z + \eta k_x  \sigma_z
\end{equation}

Finally, diagonalizing this Hamiltonian in the truncated Hilbert space spanning 
the two 
spin resolved states in the lowest subband yields the eigenenergies
\begin{equation}
E^{(3)}_{\pm} = {{\hbar^2 k_x^2}\over{2 m^*}}  + E'_0 \pm 
\sqrt{\left ( \eta^2 + \alpha_3^2 \right ) k_x^2 +  2 (\alpha_3 \Delta - \eta 
\beta) k_x + \Delta^2 +
\beta^2} 
\label{eigenenergy2}
\end{equation}
and the corresponding eigenstates
\begin{eqnarray}
{\Psi^{(3)}}_{+}(B, x) =
\left [ \begin{array}{c}
             cos(\theta''_{k_x})\\
             sin(\theta''_{k_x}) \\
             \end{array}   \right ]
             e^{i  k_x x}
             ~~~~~~~~~
{\Psi^{(3)}}_{-}(B, x) =
 \left [ \begin{array}{c}
              sin(\theta''_{k_x})\\
              - cos(\theta''_{k_x})\\
             \end{array}   \right ]
             e^{i  k_x x}
\label{eigenstate2}
\end{eqnarray}
where $\theta''_{k_x}$ = $(1/2) arctan [(\alpha_3 k_x + \Delta)/(\eta k_x - 
\beta)]$. 

In Fig. 4, we plot the dispersion relations for the case $\eta$ = $\alpha_3$ and 
$\Delta$ = $\beta$. As before, the dispersions are clearly nonparabolic and 
unless $\alpha_3 \Delta$ = $\eta \beta$,  they are asymmetric about the energy 
axis. The eigenspinors are functions 
of $k_x$ because $\theta''_{k_x}$ depends on $k_x$. Therefore, the eigenspinors 
are again not fixed in any subband, but change with $k_x$. Neither subband has a 
definite spin quantization axis and the spin of an electron in either subband 
depends 
on the wavevector. Consequently, it is again always possible to find two states 
in the 
two subbands having non-orthogonal spins and a non-magnetic scatterer  can 
couple these two states causing a spin-relaxing 
scattering event.

\begin{figure}
\epsfxsize=4.3in
\epsfysize=4.3in
\centerline{\epsffile{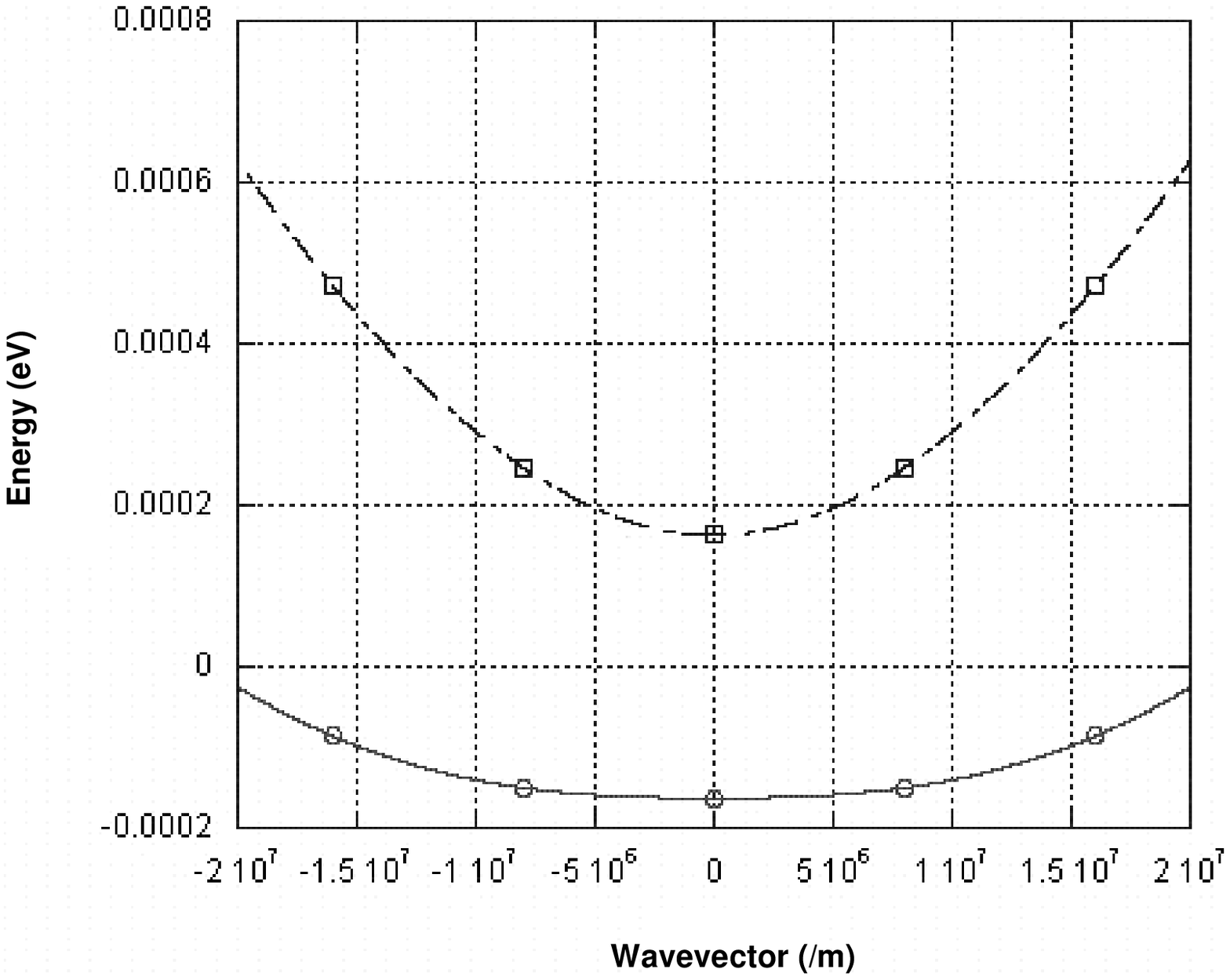}}
\caption[]{\small Energy versus wavevector $k_x$ in a quantum wire where the 
magnetic flux density is along the z-direction. We have assumed $\Delta$ = 
$\beta$ = 0.115 meV and $\alpha_3$ = $\eta$ = 10$^{-11}$ eV-m. All other 
parameters are the same as in Fig. 2. This dispersion is symmetric about the 
energy axis only because the parameters were so chosen that $\alpha_3 \Delta$ = 
$\eta \beta$; otherwise, it would have been asymmetric.}
\end{figure} 

\subsection{Spin components}

The spin components along the $x$, $y$ and $z$ directions are given by
\begin{equation}
S^{(3) \pm}_m = [\Psi^{(3)}_{\pm}]^{\dag}[\sigma_m][\Psi^{(3)}_{\pm}] ~~~ m = 
x,y,z
\end{equation}
This yields $S^{(3) \pm}_x(k_x) = \pm sin(2 \theta''_{k_x})$, $S^{(3) \pm}_y$ = 
0 and 
$S^{(3) \pm}_z (k_x) = \pm cos(2 \theta''_{k_x})$.

In this case, $S^{(3)}_x(k_x) \neq  S^{(3)}_x(-k_x)$ and $S^{(3)}_z(k_x) \neq 
S^{(3)}_z(-k_x)$ unless the Dresselhaus interaction vanishes so that $\alpha_3 = 
\Delta$ = 0. In the latter case, $S^{(3)}_x(k_x) = -  S^{(3)}_x(-k_x)$ and 
$S^{(3)}_z(k_x) = S^{(3)}_z(-k_x)$.

\section{Conclusion}

We have shown here that the dispersion relations of the lowest eigenstates in a 
quantum wire placed in a magnetic field are always non-parabolic and the 
subbands do not have fixed spin quantization axes since the eigenspinor in any 
subband changes with the wavevector. As a result,  a non-magnetic scatterer can 
always couple two states in two subbands and flip spin. However, if the magnetic 
field is absent, then the eigenspinors become wavevector-independent since then 
$\theta = \theta' = \theta'' = (1/2)arctan[\alpha_m/\eta]$  ($m$ = 1,2,3). In 
that case, each subband has a definite spin quantization axes and the 
eigenspinors in two subbands are orthogonal. Consequently, a non-magnetic 
scatter cannot couple the two subbands and flip spin. Spin transport can then be 
ballistic, particularly because the Elliot-Yafet spin flip mechanism 
\cite{elliott} will also be suppressed since the eigenspinors are wavevector 
independent. This effect, whereby a magnetic field can cause a crossover from 
ballistic to non-ballistic spin transport, can have applications in magnetic 
field sensors \cite{bandy}.



\end{document}